%% file: mesons.tex
\documentclass[aps,prd,showpacs,twocolumn,superscriptaddress,nofootinbib,preprintnumbers]{revtex4}

\usepackage{amsmath}
\usepackage{amsfonts}
\usepackage{amssymb}
\usepackage{graphicx}
\usepackage{epstopdf}
\usepackage{longtable}

\def\bfx{{\bf x}}

\begin{document}

\preprint{CP3-Origins-2010-12, WUB/10-06}

\title{Mesonic spectroscopy of Minimal Walking Technicolor}


\author{Luigi Del Debbio}
\email[]{luigi.del.debbio@ed.ac.uk}
\affiliation{SUPA, School of Physics and Astronomy, University of Edinburgh,
	Edinburgh EH9 3JZ, Scotland}
\author{Biagio Lucini}
\email[]{b.lucini@swansea.ac.uk}
\affiliation{School of Physical Sciences, Swansea University,
Singleton Park, Swansea SA2 8PP, UK}
\author{Agostino Patella}
\email[]{a.patella@swansea.ac.uk}
\affiliation{School of Physical Sciences, Swansea University,
Singleton Park, Swansea SA2 8PP, UK}
\author{Claudio Pica}
\email[]{pica@cp3.sdu.dk}
\affiliation{CP$^3$-Origins, University of Southern Denmark,
	Odense, 5230 M, Denmark}
\author{Antonio Rago}
\email[]{rago@physik.uni-wuppertal.de}
\affiliation{Department of Physics, Bergische Universit\"at Wuppertal, Gaussstr. 20, D-42119 Wuppertal, Germany}


\date{\today}

\begin{abstract}
We investigate the structure and the novel emerging features of the
mesonic non-singlet spectrum of the Minimal Walking Technicolor (MWT)
theory.  Precision measurements in the nonsinglet pseudoscalar and vector
channels are compared to the expectations for an IR-conformal field theory
and a QCD-like theory.  Our results favor a scenario in which MWT is
(almost) conformal in the infrared, while spontaneous chiral symmetry
breaking seems less plausible. 
\end{abstract}


\maketitle

\section{Introduction}
\label{sec:intro}
The idea of a new strong force in our model of Nature to explain Electro--Weak
Symmetry Breaking (EWSB) dynamically was first suggested many years
ago~\cite{Weinberg:1975gm,Susskind:1978ms}.  Borrowing from our intuition of
QCD, a new strong sector beyond the standard model was proposed in which chiral
symmetry breaks down at the TeV scale leading to EWSB and providing an
explanation for the observed gauge boson masses. Moreover, the standard model
fermions acquire their masses via extended technicolor interactions. The first models based on these
ideas were obtained by a naive rescaling of QCD, i.e. they were based on a
SU(N) gauge theory with a small number of fundamental matter fields.  Despite the elegance of the
proposal, it was soon shown that such models are not viable candidates: 
together with the mass of SM
particles, large Flavor Changing Neutral Currents (FCNC) and large values of the
Peskin-Takeuchi~\cite{Peskin:1990zt,Peskin:1991sw} parameters would also be
generated.  Electroweak precision tests performed at
LEP~\cite{Amsler:2008zzb} put tight experimental constraints on FCNC and the oblique
parameters, which are incompatible with such predictions. 

However, the naive rescaling arguments leading to the above conclusions can 
be flawed if the dynamics of the new sector is sufficiently different
from QCD. In fact the intrinsic difficulty of handling strongly interacting
models has not stopped the theoretical speculations. 
Walking and conformal technicolor theories have been
proposed~\cite{Holdom:1984sk,Yamawaki:1985zg,Appelquist:1986an,Luty:2004ye}
whose large-distance dynamics is expected to be very different from the one of QCD.
In particular it was shown that models falling in these frameworks could
satisfy the experimental constraints~(for recent reviews of
techicolor models see~\cite{Hill:2002ap,Sannino:2008ha,Sannino:2009za,Piai:2010ma}). 

Good candidate models in these frameworks are those which lie close to the lower boundary of the so-called conformal window, where the presence of an (approximate) IR fixed point 
is believed to significantly change the non-perturbative dynamics of the theory.

The use of matter fields in
higher dimensional representations 
has been recently advocated~\cite{Sannino:2004qp,Dietrich:2006cm}
as an effective and economic way of satisfying all known experimental
constraints.  

In this work we will focus on one of these candidate theories, the so-called
Minimal Walking Technicolor theory, based on the gauge group SU(2)
with two Dirac fermions in the adjoint representation.  

All viable candidate models share the common property of being
strongly interacting at the electroweak scale and as such 
are not fully under control by analytic methods only.
Although the analytical approaches are indispensable to show
which models are the most promising, they all depend on
uncontrolled approximations or conjectures based on educated
guesses. 

In this work we study MWT using first-principle numerical simulations,
which allow to investigate the full non-perturbative dynamics of the theory.
We use the same techniques matured during the last decades for Lattice QCD
and which are now source of reliable and valuable informations for the 
phenomenology of the strong interactions at high-energy experiments.
Within the Lattice Gauge Theory framework quantitative
predictions can be obtained, which demonstrate if a candidate model is 
indeed viable or not.

In the last two years renewed interest among the lattice community has 
led to an increasing number of studies by several different
groups~\cite{Catterall:2007yx,Appelquist:2007hu,DelDebbio:2008wb,Shamir:2008pb,Deuzeman:2008sc,DelDebbio:2008zf,Catterall:2008qk,Svetitsky:2008bw,DeGrand:2008dh,Fodor:2008hm,Fodor:2008hn,Deuzeman:2008pf,Deuzeman:2008da,Hietanen:2008vc,Jin:2008rc,DelDebbio:2008tv,DeGrand:2008kx,Fleming:2008gy,Hietanen:2008mr,Appelquist:2009ty,Hietanen:2009az,Deuzeman:2009mh,Fodor:2009nh,DeGrand:2009mt,DeGrand:2009et,Hasenfratz:2009ea,DelDebbio:2009fd,Fodor:2009wk,Fodor:2009ar,Appelquist:2009ka,DeGrand:2009hu,Catterall:2009sb,Bursa:2009we,Lucini:2009an,Pallante:2009hu,Bilgici:2009kh,Machtey:2009wu,Moraitis:2009xt,Kogut:2010cz,Hasenfratz:2010fi}.

In this work we present a detailed study of the non-singlet mesonic sector
of the spectrum of the gauge theory SU(2) with two Dirac adjoint fermions.
In a companion paper \cite{DelDebbio:2010xxx} we will present our result for the
glueball mass spectrum and string tension and compared them to the one
obtained in this paper for the mesonic spectrum as first suggested in
\cite{DelDebbio:2009fd}. As the simplest of such interesting models, it is
particularly amenable to numerical investigations.

Given the present analytical uncertainties, it is not clear if this theory
lies within the conformal window or not.  To understand if this model lies
within the conformal window, in this paper we will compare our data to the
signatures of spontaneous chiral symmetry breaking on the one hand and to
the expected scaling behavior in proximity of an IR fixed point on the
other.  By studying the dependence of the low-lying meson masses on the
current quark mass we will provide evidence for the existence of an IR
fixed point.

This paper is organized as follows. 
In Sect.~\ref{sec:RGflow} we remind to the reader
the physical implications of the existence of an IR fixed point
in the theory and its observable consequences as derived from a
Renormalization Group (RG) analysis of the fixed point.
In Sect.~\ref{sec:sim} we introduce the Lattice Gauge Theory formalism used 
this work.
In Sect.~\ref{sec:res} we present our numerical results and compare them to the theoretical expectations.
We finally conclude in Sect.~\ref{sec:concl}.

\section{Non-QCD behavior}
\label{sec:RGflow}

Strongly interacting theories that have a dynamics different from QCD
are needed in order to be able to build successful phenomenological
models of DEWSB. One example of such theories is provided by gauge
theories in the so-called {\em conformal window}, which are
characterized by the existence of an infrared fixed point in their
renormalization group flow.

A four--dimensional gauge theory minimally coupled to fermions in some
representation of the gauge group has a perturbative UV fixed point
provided the number of fermion species is not too large; at the UV
fixed point the gauge coupling $g$ and the fermion mass $m$ are
relevant couplings. The gauge coupling $g$ is dimensionless in 
four dimensions. It is convenient to use dimensionless couplings for
discussing RG flows. Hence we shall consider the dimensionless
quantity $\hat m=am=m/\mu$ when studying the RG transformations of the
fermion mass. Renormalized trajectories, {\it i.e.} lines of constant
physics, are one--dimensional curves originating from the UV fixed
point. Points on a given renormalized trajectory correspond to
theories that have the same long-distance physics, but different
values of the UV cutoff. Each line corresponds to a theory with a
given physical fermion mass.

One of the lines of constant physics corresponds to the massless
renormalized trajectory. If the theory possesses an IRFP, the latter
has to lie on the massless trajectory, otherwise the finite fermion
mass would drive the theory away from the IRFP at large
distances. Assuming the existence of such a fixed point, we can
linearize the RG equations in the vicinity of the fixed point, and
identify relevant and irrelevant directions. In particular the mass
$\hat m$ will be a relevant operator.

The running of the couplings is described by the RG equations:
\begin{eqnarray}
  \label{eq:gbeta}
  \mu \frac{d}{d\mu}g &=& \beta(g)\, ,\\
  \label{eq:mbeta}
  \mu \frac{d}{d\mu}m &=&  -\gamma(g) m\, ,
\end{eqnarray}
where $g$ and $m$ are respectively the running coupling and the
running mass, which depend on the energy scale $\mu$. Note that chiral
symmetry guarantees that the RHS of Eq.~(\ref{eq:mbeta}) is
proportional to the mass itself. The function $\gamma$ is the anomalous
dimension of the scalar density operator $\bar\psi(x)
\psi(x)$. Eq.~(\ref{eq:mbeta}) implies:
\begin{equation}
  \label{eq:hbeta}
  \mu \frac{d}{d\mu}\left(\frac{m}{\mu}\right) =  -\left[\gamma+1\right] 
  \left(\frac{m}{\mu}\right)\, .
\end{equation}

Similar equations describe the evolution of all the other couplings
that are compatible with the symmetries of the system under study. We
shall denote the generic, dimensionless coupling $\hat g_i$; their
evolution is dictated by a corresponding $\beta$ function:
\begin{equation}
  \label{eq:betasystem}
  \mu \frac{d}{d\mu}\hat g_i = \beta_i(\hat g)\, .
\end{equation}

The IRFP is defined by an isolated zero of the $\beta$ functions.
Theories in the conformal window become scale--invariant at large
distances, and therefore cannot develop condensates. In particular
chiral symmetry cannot be spontaneously broken, and there are no
single--particle states; the dynamics is entirely expressed by the
exponents that characterize the power--law behavior of field
correlators at large distances.

In the vicinity of a fixed point $\hat g^*$ the RG equations can be
linearized; the evolution is characterized by the matrix:
\begin{equation}
  \label{eq:linRG}
  R_{ij} = \left.\frac{\partial \beta_i}{\partial \hat
      g_j}\right|_{\hat g^*}\, .
\end{equation}
The evolution of the dimensionless eigenvectors of the matrix $R$,
$u_i$, is given by simple power laws:
\begin{equation}
  \label{eq:RGeig}
  u_i(\mu) \propto \mu^{-y_i}\, ,
\end{equation}
where $y_i$ are the eigenvalues of $R_{ij}$. The $y_i$ are the
critical exponents that are commonly used in the theory of critical
phenomena. It is clear from Eq.~(\ref{eq:RGeig}) that $y_i>0$
characterizes the relevant directions at the IRFP.
 
The fermion mass is a relevant operator at the fixed point, and we can
readily deduce from Eq.~(\ref{eq:hbeta}):
\begin{equation}
  \label{eq:ym}
  y_m=\gamma_*+1\, ,
\end{equation}
where $\gamma_*$ is the value of the anomalous dimension at the fixed
point. 

The scaling (or conformal) dimension of the scalar density $\Delta_m$
is related to the critical exponent $y_m$ by:
\begin{equation}
  \label{eq:Deltam}
  y_m = D -\Delta_m\, ,
\end{equation}
where $D$ is the dimension of space-time. The scaling dimension for a
scalar operator is bound to be greater than one by unitarity, and it
is equal to three for the scalar density in the free theory. This
corresponds to the usual range $0\leq \gamma_* \leq 2$.

\paragraph*{\bf Scaling of the free energy density.}

Let us now consider a RG transformation such that the lengths are rescaled
by a factor $b$:
\begin{equation}
  \label{eq:bfactor}
  a^\prime = b a\, ,~~~~ \mu^\prime=\mu/b\, .
\end{equation}
Following the discussion above, the transformation of the singular part of
the free energy density under such transformation can be written as:
\begin{equation}
  \label{eq:RGfs1}
  f_s(u_i,am,a/L) = b^{-D} f_s(b^{y_i} u_i,b^{y_m} am,b a/L)\, .
\end{equation}

We have denoted by $u_i$ the irrelevant operators at the IRFP, and
therefore $y_i<0$, while $y_m>0$ as expected from the discussion above on
the role of the fermion mass.  We have included the dependence on the size
of the box $L$, since this will provide finite-size scaling laws. The
inverse of the size $1/L$ is treated as relevant coupling with unit
eigenvalue; the underlying hypothesis here is that the finite value of $L$
does not affect the RGE for the other couplings, {\it i.e.} that it is
larger than the inverse mass of the states in the theory, $L m \ll 1$.

Iterating the RG transformation $n$ times yields:
\begin{equation}
  \label{eq:RGfs2}
  f_s(u_i,am,a/L) = b^{-nD} f_s(b^{ny_i} u_i,b^{ny_m} am,b^n a/L)\, .
\end{equation}
Choosing $n$ such that $b^{ny_m} am=am_0$, where $m_0$ is some
reference mass scale, Eq.~(\ref{eq:RGfs2}) can be rewritten as:
\begin{widetext}
\begin{eqnarray}
  \label{eq:RGfs3}
  f_s(u_i,am,a/L) = \left(\frac{m}{m_0}\right)^{\frac{D}{y_m}}
  f_s\left(\left(\frac{m}{m_0}\right)^{-\frac{y_i}{y_m}}u_i,am_0,
  \left(\frac{m}{m_0}\right)^{-\frac1{y_m}}a/L\right)
  = \left(\frac{m}{m_0}\right)^{\frac{D}{y_m}}
  \Phi\left(mL^{y_m},
  \left(\frac{m}{m_0}\right)^{\frac{|y_i|}{y_m}}u_i\right)\, .
\end{eqnarray}
\end{widetext}
Eq.~(\ref{eq:RGfs3}) describes the scaling with the fermion
mass, and the functional form of finite-size effects. Expanding $\Phi$
in powers of $x_i=\left(\frac{m}{m_0}\right)^{|y_i|/y_m}u_i$,
assuming as usual that $f_s$ is analytic as a function of the irrelevant
couplings, yields the corrections to the scaling. 
Note that these corrections vanish when $x=0$.

Ignoring the corrections to scaling we obtain an expression for the
free energy that is useful to derive finite-size scaling properties:
\begin{equation}
  \label{eq:RGfs4}
  f_s(am,a/L) = \left(\frac{m}{m_0}\right)^{D/y_m} 
  \Phi(mL^{y_m})\, .
\end{equation}

\paragraph*{\bf Scaling of correlators.} 

Two--point function correlators:
\begin{equation}
  \label{eq:Hcorr}
  f_H(x;am)=\int d^{D-1}x \langle H(x) H(0)^\dagger\rangle
\end{equation}
satisfy similar RG equations:
\begin{equation}
  \label{eq:HcorrRGE}
  \left[ a\frac{\partial}{\partial a} - \gamma m
    \frac{\partial}{\partial m} - 2 \gamma_H\right]f_H(x;am)=0\, ,
\end{equation}
where we have neglected the dependence on the irrelevant
couplings, and $\gamma_H$ is the anomalous dimension of the 
field $H$. The solution to the above equation obeys the following
scaling law:
\begin{equation}
  \label{eq:RGcorr1}
  f_H(x;am) = b^{-2\gamma_H} G(x;b^{y_m} am)\, ,
\end{equation}
Iterating this relation $n$ times, and performing manipulations
analogous to the ones above, yields:
\begin{equation}
  \label{eq:RGcorr2}
    f_H(x;am) = \left(\frac{m}{m_0}\right)^{2\gamma_H/y_m} 
    \Psi\left(x/|m/m_0|^{-1/y_m}\right)\, .
\end{equation}
If the correlator decays exponentially at large distances with the
mass of the lightest state that overlap with the fields, then
Eq.~(\ref{eq:RGcorr2}) determines the scaling of this mass:
\begin{equation}
  \label{eq:Mscal}
  M \sim m^{1/y_m}\, .
\end{equation}
Using the relation between $y_m$ and $\gamma_*$ discussed above, the
scaling for the masses of the physical states is
\begin{equation}
  \label{eq:massscal2}
    M \sim m^{1/(1+\gamma_*)}\, .
\end{equation}
Note that according to this analysis all masses scale with the same
exponent.  Different states are selected by different fields appearing
in the correlator $f_H$, which implies that the anomalous dimension in
the prefactor, $y_H$, does change. However the scaling of the mass is
entirely due to the RG behavior of the argument of the function
$\Psi$, which does not depend on the state under scrutiny.

\section{Lattice simulations}
\label{sec:sim}
Non-perturbative numerical simulations are performed after introducing an
effective UV and IR cutoff in the form of a space-time lattice of
finite extent. 

The Euclidean path integral is thus reduced to an ordinary integral over a
large number of degrees of freedom. 
The choice of a discretized action on the lattice is not unique;
different choices will result in different lattice artefacts.
Unimproved Wilson fermions are used throughout this work.  The lattice
action with matter fields in a representation $R$ is given
by\footnote{We omit for the sake of simplicity all the position, color
  and spin indexes.}
\begin{equation}
  S(U, \psi, \overline{\psi})= S_g(U) + 
   \sum_{i=1}^{n_f}\overline{\psi_i} D_m(U^R) \psi_i\,\, ,
\end{equation}
with explicit expressions for the gauge action $S_g$ and the massive lattice
Dirac operator $D_m$ given below.  For this choice of discretization, the
action depends only on two bare parameters: the bare inverse coupling $\beta$
and the bare dimensionless quark mass $a m_0$, $a$ being the lattice spacing.
While the link variables appearing in the gauge
action are in the fundamental representation\footnote{The link variables can
also be taken in a different representation of the gauge group if one chooses
to.} of the gauge group, the links in the lattice Dirac operator are in the
same representation $R$ as the fermion fields.  The partition function, after
integrating out the matter fields, takes the form:
\begin{equation}
  Z = \int \text{exp}[- S_g(U)]  [{\rm det}\ D_m(U^R)]^{n_f} dU \,\, .
\end{equation}

For the Wilson action used in this work the gauge action is given by
\begin{equation}
S_g(U) = \frac{\beta}{N_c} \sum_{x, \mu<\nu} \mathrm{Re}\ \mathrm{tr}\
\mathcal{P}_{\mu\nu}(x)\,\, , 
\end{equation}
where $\mathcal{P}_{\mu\nu}(x)$ is the elementary 1$\times$1
plaquette in the $\mu-\nu$ plane at lattice site $x$.
In the fermion action the Wilson--Dirac operator is given by
\begin{equation}
  D_m(U^R) = a m_0 + \frac12 \sum_\mu \left[\gamma_\mu \left(\nabla_\mu + \nabla^*_\mu \right) 
    - a \nabla^*_\mu \nabla_\mu \right]\, ,
\end{equation}
where $\nabla_\mu$ is the discretized forward covariant derivative
depending on the link $U^R_\mu$ and $\nabla^*_\mu$ its adjoint operator:
\begin{equation}
(\nabla_\mu \psi) (x) = U^R(x,\mu)\psi(x+\mu)-\psi(x)\,\, .
\end{equation}

In a numerical lattice simulation the computation of the discretized
path integral is performed by Monte Carlo integration using
importance sampling: an \textit{ensemble} of gauge configurations is
generated with probability proportional to $\exp[-\beta S_g(U)] [\det\
D_m(U^R)]^{n_f}$.  The expectation value of any observable can then be computed
as a stochastic average over this ensemble of configurations.

\subsection{Sources of systematic errors}
In order to obtain continuum values for the observables of a theory from
numerical simulations of its lattice discretised version, an appropriate
limiting procedure must be performed. It
is thus important to understand when and to what extent the outcome of lattice
simulations are a faithful depiction of the continuum physics.  This is
especially important when one tries to understand a new theory as the MWT in
the present work, since we lack the insight and the experimental input we have
for instance in the more familiar case of QCD. In fact, in order to be the
description of a new force of Nature, this theory has to be rather different
from QCD and we need to ensure that we are observing genuine features of the
continuum theory, and not artefacts of our lattice formulation. 

We will now list the most important sources of systematic errors which are
present in lattice simulations and what are the appropriate limits to take in order 
to recover the continuum physics. In addition to these, statistical errors are also 
always present, but those can be reduced arbitrarily by producing a big enough ensemble
of configurations. 

\paragraph*{\bf Finite-size, finite-temperature effects.}
  These are due to the presence of an IR cutoff in the form of a finite extent
of the 4-dimensional lattice both in the spacial and temporal directions. The
standard lattice geometry used in numerical simulations is $T$$\times$$L^3$,
i.e. the three spacial directions have equal length.
The correct vacuum expectation
values of the continuum theory are recovered in the limit in which
$T,L\rightarrow\infty$.  On a 4-dimensional torus these expectation values can
have large corrections, even if  asymptotically the infinite volume limit is
reached at an exponentially fast rate.
As the system is tuned closer to a critical point, the magnitude of finite-size
effects and the autocorrelation of lattice observables increase.
This is what happens for example when particles with a Compton wavelength
comparable with the lattice size are present.
Moreover, if the size of the lattice is not sufficiently large, the
system may enter a phase that bears little resemblance to the large--volume
theory we are interested in. Some examples, based on analytical finite--volume
results~\cite{'tHooft:1979uj,Luscher:1982ma,vanBaal:1986ag}, have been recently
discussed in Ref.~\cite{Fodor:2009wk}.
\paragraph*{\bf Explicit breaking of chiral symmetry.}
It is difficult and numerically very expensive to preserve chiral symmetry once
the theory is discretized on a lattice. For the particular choice of the
lattice action used in the work, i.e. Wilson--Dirac fermions, chiral symmetry
is explicitly broken at the Lagrangian level.
In order to recover it in the continuum limit the tuning
of one parameter (the bare quark mass) is necessary. 
Even with lattice chiral fermions,
  the low modes of the lattice Dirac operator, which appear as the
  chiral limit is approached, make it practically impossible to run
  simulations at very light masses with the current algorithms.
When using Wilson fermions on a given finite lattice, there is a lower limit to
the masses that are numerically accessible. This limit depends on the volume
of the system~\cite{DelDebbio:2005qa,DelDebbio:2007pz}.
\paragraph*{\bf Discretization artefacts.}
The space-time lattice introduces a UV cutoff, i.e. the lattice spacing
$a$. The interesting continuum physics is recovered in
the limit where the lattice spacing goes to zero.  However since all the
quantities in a simulation are dimensionless, the lattice spacing $a$ is not a
parameter that one can directly change. Instead the value of the lattice
spacing is inferred from the measure of a physical quantity on the lattice,
which is thus used to fix the scale $a$. The continuum limit is recovered when
the system is tuned to a UV critical point where the lattice spacing
vanishes. This can be done in the space of bare parameters by increasing the
lattice inverse coupling $\beta$ to infinity, i.e. sending the coupling of the
theory $g$ to zero, as dictated by asymptotic freedom.  At that point the
system undergoes a continuous phase transition and, by universality, the
microscopic details of the system become immaterial. To put it in a different
way, the physics at the scale of the lattice spacing must decouple from the
long range (continuum) physics, which is what we are interested in studying
on the lattice.

To give reliable predictions, lattice simulations must be performed in a regime
in which a precise hierarchy of scales is realized. To illustrate this point,
let us consider first the case of QCD.  Numerical simulations, as explained
above, are always performed with a finite quark mass, which explicitly induces
a mass gap in the theory. To avoid finite-size effects in the computation of
the low-lying spectrum, the lattice size $L$ must be much bigger than the
inverse mass of the lightest hadron $M_\mathrm{PS}$ we aim to measure.
If we are interested in the chiral regime, as it is usually the case, this light
hadron mass must be much smaller than the characteristic hadronic scale at
which the theory becomes strongly interacting. In QCD we can take for example
the Sommer scale $r_0$ as a convenient quantity to measure on the lattice.
Finally to avoid discretization errors, the lattice spacing must be much
smaller than the reference scale for strong interactions $r_0$, so that the
physics at the scale of the UV cutoff is weakly interacting. In summary we
must have the following hierarchy of scales:
\begin{equation}
  L^{-1} \ll M_\mathrm{PS} \ll r_0^{-1} \ll a^{-1}\,\, ,
\end{equation} 
for the computations to reproduce reliably the features of the continuum
chiral regime of QCD.

Let us now consider the case in which the underlying continuum theory has an IR
fixed point in the massless limit. The presence of a mass term in numerical
simulations explicitly breaks conformal invariance and as in the QCD-like case
a mass gap and a particle spectrum for the theory are generated.
In order to reliably estimate a hadron mass, we still need the lattice to be
big enough to fit the state
we are interested to measure, i.e. as above $L^{-1} \ll M_\mathrm{PS}$, for
example.  However in contrast to the QCD-like case, all the masses of the
hadrons vanish as we send the explicit symmetry breaking term to zero. 
One can still define an IR scale like $r_0$ but, as the theory is no longer
confining, $r_0$ is no longer related to the particle masses
and it is therefore not a useful quantity to compare to. One can be tempted
however to introduce a modified version of the IR scale $\Lambda$ which marks
the onset of the IR scaling region. With this definition the interesting mass
region to explore is $M_\mathrm{PS} \ll \Lambda$, in analogy with the previous
case.  Finally also in this case discretization artefacts should be suppressed using
fine lattices.  In summary we have a similar hierarchy of scales as before
\begin{equation} 
L^{-1} \ll M_\mathrm{PS} \ll \Lambda \ll a^{-1}\,\, ,
\end{equation} 
with a new IR scale $\Lambda$ whose definition is related to the
new features of the theory. Unfortunately a convenient definition of $\Lambda$
readily measurable on the lattice, like the Sommer scale for QCD, is not at hand.
As defined above $\Lambda$ can only be inferred \textit{a posteriori} by observing
the scaling behavior of some physical observable.

\subsection{Simulation Code}
The results presented in this work are obtained using our own simulation code,
written from scratch for the specific purpose of studying gauge theories with
fermions in arbitrary representations.  The code was developed during the last
few years and has been presented and tested in detail in
Ref.~\cite{DelDebbio:2008zf}.  
This code was designed to be flexible and
easily accommodate for fermions in arbitrary representations of the gauge
group, without compromizing the performance and ease of use of the code itself.

Our simulation code, named HiRep, is suitable for the study of gauge theories with
\begin{itemize}
\item gauge group $SU(N)$ for any $N$. The code has already been used for the study of
the large-N mesonic spectrum~\cite{DelDebbio:2007wk} with $N$ up to 6.
\item generic fermion representations. At present the code implements
  the fundamental (fund), adjoint (ADJ), 2-index symmetric (S) and
  antisymmetric (AS). All of these different representation have already
been successfully tested and used~\cite{Armoni:2008nq}. It is easy to extend the code to other
  representations, like 3-index symmetric for example, and even to
  have fermions in two or more different representations at the same
  time. In particular no modifications for the computation of the HMC
  force are needed, which is typically the most complicated part of the
  code.
\item any number of flavors. We use Wilson fermions with the
HMC\cite{Duane:1987de}/RHMC\cite{Kennedy:1998cu,Clark:2003na}
  algorithm, which is an exact algorithm for any number of flavors.
\end{itemize}
We have also implemented a significant number of observables such as the 
the measure of the mesonic spectrum (presented in this work), 
Schr\"odinger functional observables (used in Ref.~\cite{Bursa:2009we}),
gluonic observables like Wilson and Polyakov loops, glueball masses
presented in a companion paper~\cite{DelDebbio:2010xxx}.
 
Since this was a new code, and lattice simulations in the past were
mainly devoted only to QCD, we made a large effort to validate the
code and to study its behavior in the parameter region of interest for
the physics. As part of this effort, in particular we made a number of
cross-checks:
\begin{itemize}  
\item for $SU(3)$, $n_f=2$ with different, established codes (we used
  M.~Luscher's DD-HMC algorithm~\cite{Luscher:2005rx});
\item consistency among different representations (e.g. $SU(3)$ AS vs
  fund, $SU(2)$ SYM vs ADJ);
\item observables with different codes (e.g. meson spectrum for
  $SU(2)$ ADJ in Chroma);
\item large quark mass limit compared to pure gauge (quenched)
  spectrum;
\item correctness of integrator, reversibility, acceptance
  probability;
\item independence from integrator step size.
\end{itemize}
To control the stability of the simulations, which could incur in well-known
problems close to the chiral limit~\cite{DelDebbio:2005qa}, we monitored the
lowest eigenvalue of the (pre-conditioned) Wilson--Dirac matrix. The average
value of the lowest eigenvalue together with the standard deviation of their
distribution can be found in the tables in Appendix~\ref{sec:tables}.  We
found no instabilities in our runs. 

\subsection{Meson masses}
Let $\Gamma$ and $\Gamma^\prime$ be two generic matrices in the Clifford
algebra, we define the two--point correlator at zero
momentum as follows:
\begin{equation}
f_{\Gamma\Gamma^\prime}(t) = \sum_\bfx \langle (\bar\psi_1(\bfx,t) \Gamma
\psi_2(\bfx,t))^\dagger \, \bar\psi_1(0) \Gamma^\prime \psi_2(0) \rangle\, ,
\end{equation}
where $\psi_1$ and $\psi_2$ represent two different flavors of
degenerate fermion fields, so that we only consider flavor
non--singlet bilinears.  Denoting the space--time position $(\bfx,t)$
by $x$ and performing the Wick contractions yields:
\begin{equation}
f_{\Gamma\Gamma^\prime}(t) = \sum_\bfx 
- \mathrm{tr} \left[ \gamma_0 \Gamma^\dagger \gamma_0 S(x,0) \Gamma^\prime S(0,x) \right] 
\, ,
\end{equation}
where $S$ denotes the quark propagator, i.e. the inverse of the hermitean Wilson--Dirac matrix $\gamma_5 D$.
In practice not all matrix elements of $S$ are computed, but only some single rows (point-to-all propagator) by solving the linear system:
\begin{equation}
D(x,y)_{AB} \eta^{\bar A,0}_B(y) = \delta_{A,\bar A} \delta_{x,0}\, ,
\end{equation}
where capital Latin letters like $A=\{a,\alpha\}$ are collective
indices for color and spin, and $\bar A$, $x=0$ is the position of
the source for the inverter. The inversion is performed using a QMR
recursive algorithm with even--odd preconditioning of the Dirac
operator, which is stopped when the residue is less than $10^{-8}$.
For some of our lattices we used the noise-reduction technique described in 
Ref.~\cite{Boyle:2008rh} to take a stocastic average over the volume of the 
point source.

Following Ref.~\cite{DelDebbio:2007pz}, masses and decay constants for
the pseudoscalar meson are extracted from the asymptotic behavior of
the correlators $f_\mathrm{PP}$ and $f_\mathrm{AP}$ at large Euclidean
time. The pseudoscalar mass and the vacuum--to--meson matrix element
are obtained from the correlator of two pseudoscalar densities:
\begin{equation}
  \label{eq:PPasym}
  f_\mathrm{PP}(t) = -\frac{G^2_\mathrm{PS}}{M_\mathrm{PS}}
  \exp\left[-M_\mathrm{PS} t\right] + \ldots\ .
\end{equation}
The meson mass is obtained by fitting the effective mass to a
constant, while the coupling $G_\mathrm{PS}$ is extracted from the
amplitude of the two--point function $f_\mathrm{PP}$. 
The definition of the effective mass used in this work is given in Appendix~\ref{sec:effmass}. 

As in Ref.~\cite{DelDebbio:2007pz} the ratio
\begin{equation}
  \label{eq:PCACmass}
  m_\mathrm{eff}(t) = \frac14 \left[\left(\partial_0 + \partial_0^* \right)
    f_\mathrm{AP}(t)\right] / f_\mathrm{PP}(t)
\end{equation}
yields the PCAC mass $m$ with corrections of $\mathcal O(a)$ for the
unimproved theory. Note that the decay constant is not computed directly; it is obtained
from the values computed above as:
\begin{equation}
  \label{eq:FPSeq}
  F_\mathrm{PS}=\frac{m}{M_\mathrm{PS}^2} G_\mathrm{PS}.
\end{equation}
The decay constant extracted from bare lattice correlators is related
to its continuum counterpart by the renormalization constant $Z_A$,
which has been computed in perturbation theory in
Ref.~\cite{DelDebbio:2008wb}.

Finally the mass of the vector state is extracted from the $f_\mathrm{VV}$
correlator, again using a fit to the effective mass plateaux.

On the smallest lattices that we have used in this study, it is difficult to
isolate clearly the contribution from the lowest state, which dominates the
large--time behavior of two--point correlators; this yelds large systematic
errors. We have however explicitly measured the meson masses at the same value
of the bare parameters -- corresponding to the same value of the PCAC mass -- on
increasingly larger lattices to ensure that the residual finite-volume
corections on these quantities are small. 

For the pseudoscalar and vector channels we have also verified that different
choices of interpolating operators give results that agree with each other.
In these channels
the final results presented below are obtained taking the average over the
different choices of interpolating operators.

\subsection{Simulation parameters}
All the simulations discussed in this work are performed
at a fixed lattice spacing,
corresponding to a bare coupling $\beta=2.25$. This value of the coupling was
chosen based on previous studies of the same
theory~\cite{Catterall:2007yx,Catterall:2008qk,Hietanen:2008mr} to avoid a bulk
phase transition present at about $\beta=2.0$.  The extrapolation towards the
continuum limit requires a new series of runs at different values of the bare
coupling, and is left to future investigations.
 
We use four different lattices: $16\times 8^3$, $24\times 12^3$, $32\times
16^3$ and $64\times 24^3$.  For each of these four lattices a number of
ensembles corresponding to different quark masses were generated, focusing in
particular on
the range corresponding to pseudoscalar masses between 0.6$a^{-1}$ and 0.2$a^{-1}$. As
explained above, when using Wilson fermions, the chiral and infinite volume
limits are intertwined. In practice for the simulation to be stable, one cannot
arbitrarily decrease the quark mass without also increasing the volume. This is
also necessary to keep under control finite-volume effects, thus remaining in
the large volume limit.  We explicitly control the size of these systematic
errors performing simulations with different volumes. 

For each lattice and quark mass we accumulated a statistical ensemble of about
5000 thermalized configurations, except at the largest volume for which we
present only preliminary data based on approximately 500 configurations for
each quark mass. The gauge configurations were generated using trajectories in
the molecular dynamics integration of length 1 for the two smallest volumes and
1.5 for the two largest lattices, with integration parameters leading to an
acceptance rate of about 85\% in all cases, and to an integrated
autocorrelation times for the lowest eigenvalue of the Wilson--Dirac operator
of order 15 or less.

Details of simulation parameters and results are reported in the tables of
Appendix~\ref{sec:tables}.

\section{SU(2) with 2 Adjoint fermions}
\label{sec:res}
Before looking at the actual numerical results, let us discuss the signatures
of an IR fixed point, in order to focus on the important aspects of our
numerical evidence.  In this work, we search for indications of IR conformal
behavior in the spectrum. This is not the only possible way, as one can, for
example, study the non-perturbative running of a coupling defined in some
particular scheme. In fact most of the claims of the existence of IR fixed
points so far have been made by looking at the evolution of the coupling in the
Schr\"odinger functional
scheme~\cite{Appelquist:2007hu,Shamir:2008pb,Hietanen:2009az,Bursa:2009we}.
However at present these studies still lack a
reliable continuum extrapolation and the claims of the existence of IR fixed
points should therefore be confirmed by more solid numerical investigations. 

If there is no IR fixed point, the theory is expected to be confining and
chiral symmetry to be spontaneously broken. We will refer to this case as
QCD-like: in the chiral limit the pseudo-scalar particles (pions) become
massless, while the other states in the spectrum remain massive.  In the small
mass regime the theory can be effectively described by a chiral lagriangian and
the familiar results of QCD can be recovered.  In particular the theory has a
non-zero pseudoscalar decay constant $F_\mathrm{PS}$ and chiral condensate
$\langle\overline\psi\psi\rangle$.  Using the PCAC mass from the axial Ward
identity, $m$, to parametrize the explicit breaking of chiral symmetry, we
expect the usual scaling $M^2_\mathrm{PS}\propto m$, as $m\rightarrow 0$. The
Gell-Mann--Oakes--Renner relation is satisfied and can be used to extract the
chiral condensate: $(M_\mathrm{PS} F_\mathrm{PS})^2/m \rightarrow
-\langle\overline\psi\psi\rangle$.

On the other hand, if the theory has an IR fixed point, the arguments of
Sect.~\ref{sec:RGflow} apply. In the scaling region, i.e. in proximity of the
IR fixed point, dimensionful physical quantities are expected to scale  with a
power law behavior. 
The universal exponents appearing in these scaling laws are
related to the anomalous dimensions of the scaling fields as shown in
Sect.~\ref{sec:RGflow}. In particular the scaling of the hadron masses is
governed by the anomalous dimension of the mass $\gamma_*$ at the fixed point.
If we parametrize the
explicit breaking of chiral and conformal symmetry by $m$, in the massless limit $m\rightarrow 0$
we expect that all hadron masses vanish proportionally to the same power of
$m$: $M_\mathrm{had} \sim m^{1/(1+\gamma_*)}$; in particular the ratio of the
vector to pseudoscalar meson masses remains finite:
$M_\mathrm{V}/M_\mathrm{PS}\rightarrow \text{const}<\infty$.  Also the
pseudoscalar decay constant $F_\mathrm{PS}$ and the chiral condensate
$\langle\overline\psi\psi\rangle$ are expected to vanish in the chiral limit.

The behavior just described assumes that the system is in an infinite volume
near the continuum limit.  However this is not always the
case in a lattice simulation. In particular the finite size of the
system can be seen as a relevant coupling which will drive the system
away from criticality under the renormalization group (RG) flow.  As
$m$ is decreased towards the chiral limit, great care must be taken to
control finite size effects as explained in Section~\ref{sec:sim}. We
will show below that these effects are quite big in the range of
masses and lattice volumes which are currently used in numerical
simulations.

\subsection{Spectrum}
\label{sec:spec}

For each ensemble of configurations, we measured the quark mass from
the axial Ward identity (PCAC mass) $m$, the pseudoscalar meson mass
$M_\mathrm{PS}$, the vector mass $M_\mathrm{V}$ and the pseudoscalar decay
constant $F_\mathrm{PS}$.

\begin{figure}
\includegraphics*[width=\columnwidth]{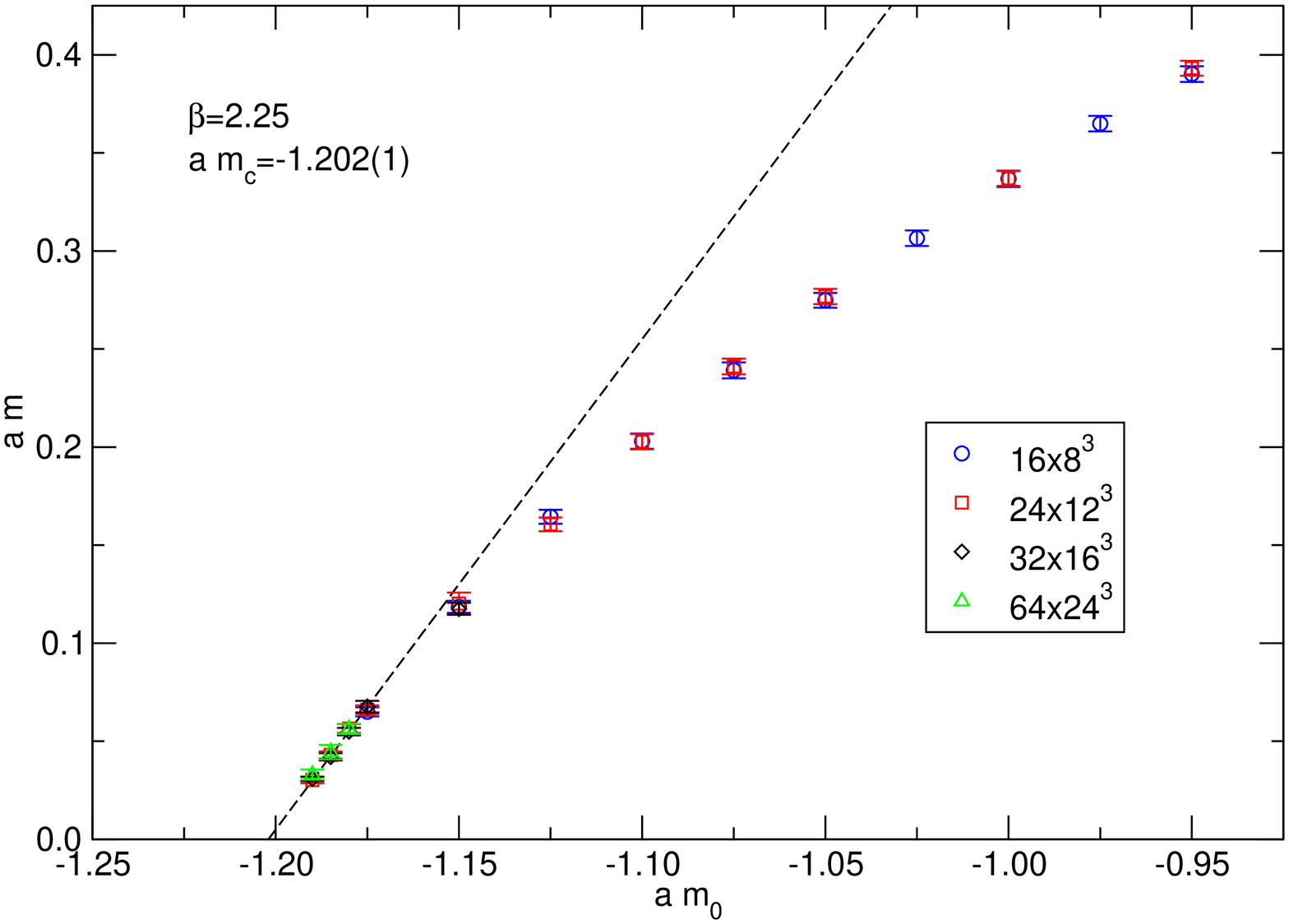}%
\caption{Extrapolation of the quark mass from the axial Ward identity
  to locate the chiral limit. As expected no significant finite size
  effects are present.\label{fig:cl}}
\end{figure}

We locate the chiral limit at the critical bare mass where the PCAC mass
vanishes. This does not correspond to zero bare mass because the explicit
breaking of chiral symmetry with Wilson fermions induces an additive
renormalization of the quark mass.  We show in Fig.~\ref{fig:cl} the
extrapolation of $m$ for different lattice sizes. Using a linear extrapolation
of the four lightest measured points, the chiral limit can be located at the
critical bare mass $am_c=-1.202(1)$. As expected from the fact that $m$ is an
UV quantity, no significant finite-size effects are visible and the measured
values for this quantity agree within errors on all four lattices.

\begin{figure*}[t]
\includegraphics*[width=\columnwidth]{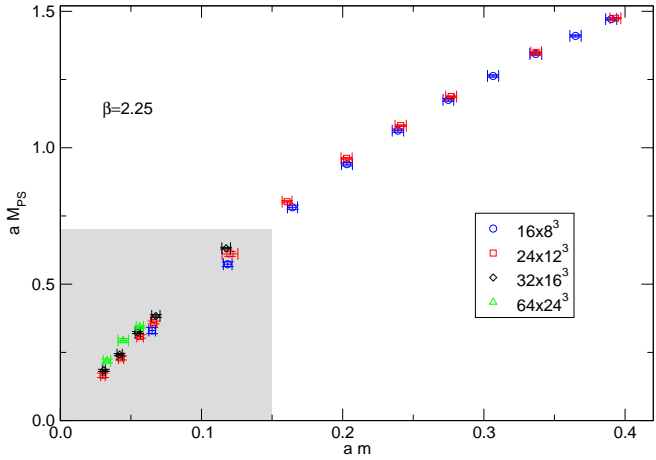}\hfill
\includegraphics*[width=\columnwidth]{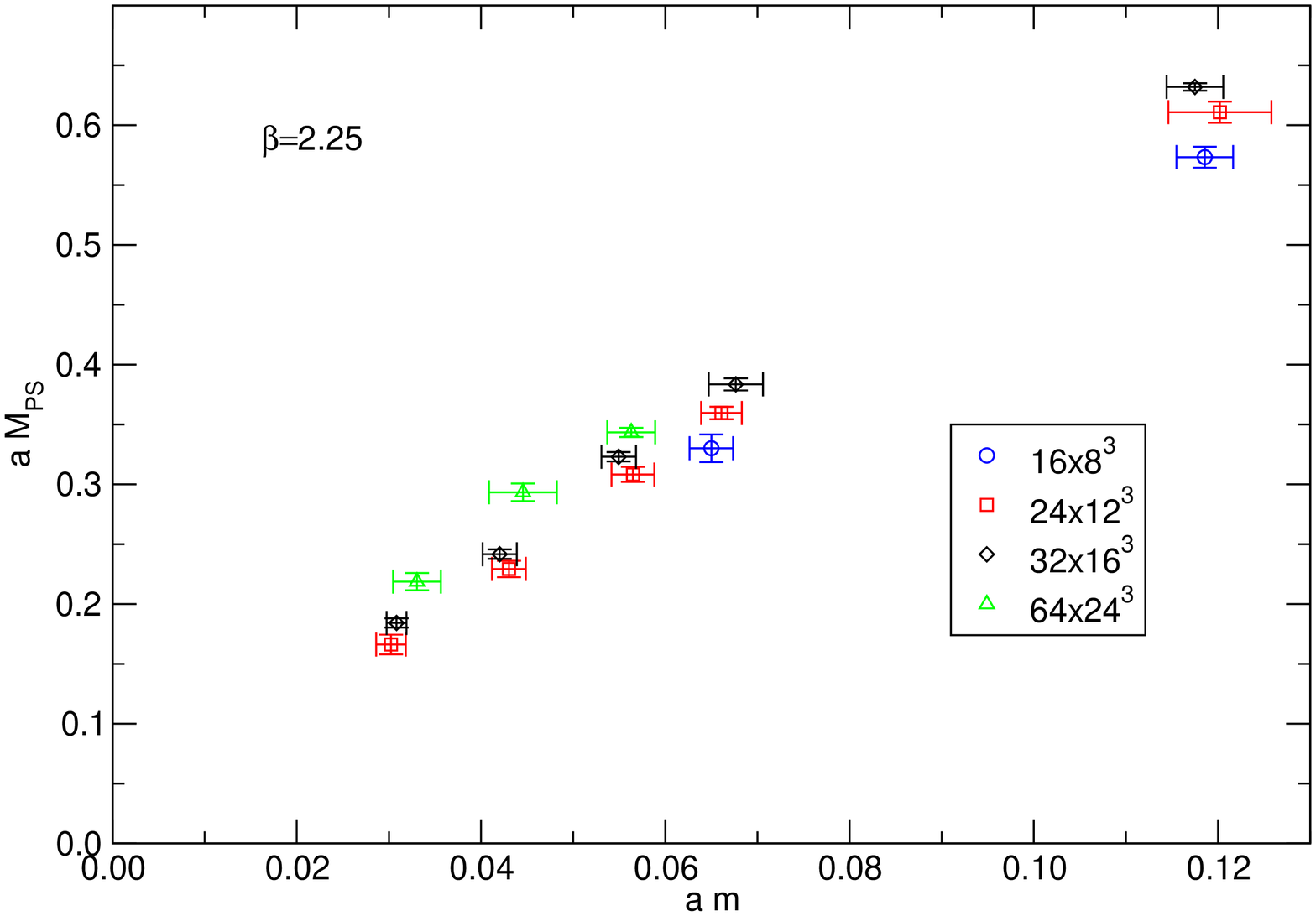}%
\caption{Pseudoscalar meson mass as a function of the PCAC mass. The
  interesting small mass region shaded in the left panel is enlarged
  on the right. Finite volume effects are evident and grow approaching
  the chiral limit.\label{fig:ps}}
\end{figure*}
Our results for the mass of the pseudoscalar meson are presented in
Fig.~\ref{fig:ps}. The interesting region of small quark masses is
shown in the right panel. 
Given the level of accuracy of the present measure, the finite volume 
systematics on $M_\mathrm{PS}$ are clearly visible and, as the PCAC
mass is decreased, they become more and more relevant, as discussed in
Sect.~\ref{sec:sim}.  To quantify this systematic effect and to keep
it under control, we use larger lattices as the chiral limit is
approached.

These large finite size effects make it harder
to draw definitive conclusions about the functional behavior of the
pseudoscalar mass in the chiral limit.
\begin{figure}
\includegraphics*[width=\columnwidth]{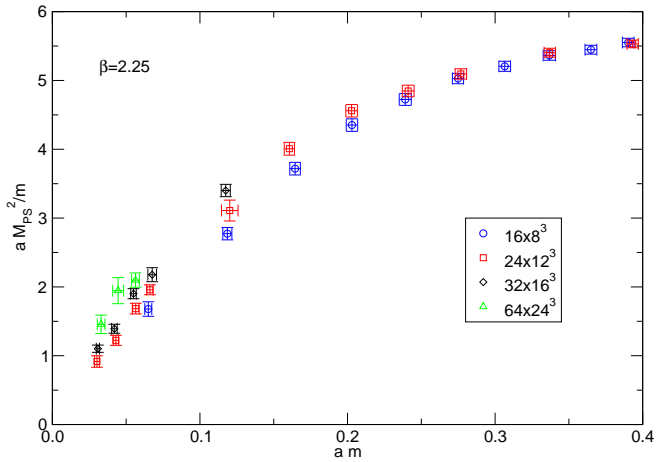}%
\caption{Ratio of the pseudoscalar mass squared to the PCAC mass. The
  extrapolation to the chiral limit suffers from large finite-volume
  effects. See the text for a discussion.\label{fig:ratio}}
\end{figure}
For a QCD-like theory the ratio $aM_\mathrm{PS}^2/m$, shown
in Fig.~\ref{fig:ratio}, should be a (non-zero) constant in the chiral limit. 
On the other hand if the theory has an IR fixed point the
ratio should vanish in the chiral limit if $\gamma_*<1$ or diverge if $\gamma_*>1$.
Our data clearly favor the IR conformal scenario with $\gamma_*<1$. 
An accurate determination of the anomalous dimension is more difficult, but from the 
almost linear behavior of $M_{PS}$ as a function of $m$ a small value of 
$\gamma_*$ seems to be preferred.

\begin{figure}
\includegraphics*[width=\columnwidth]{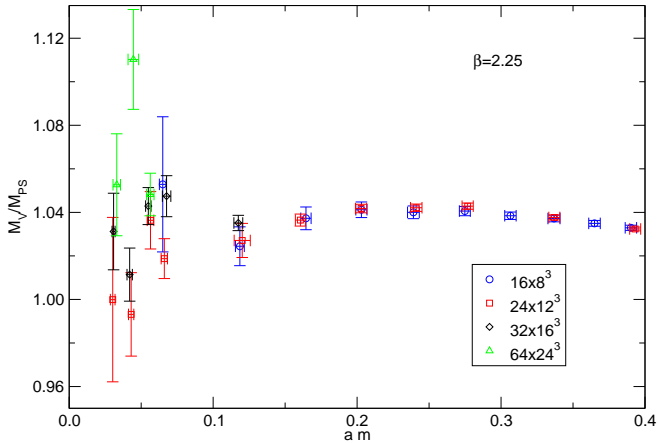}%
\caption{Comparison between the vector and pseudoscalar meson
  masses. At large PCAC mass, due to quenching the ratio is very near
  to one. Near the chiral limit large finite size effects show
  up.\label{fig:mvomps}}
\end{figure}
The IR conformal scenario is also favored when one looks at
the ratio of the vector to the pseudoscalar mass which is shown in
Fig.~\ref{fig:mvomps}. This quantity is bounded to be greater than
1~\cite{Weingarten:1983uj} and in the heavy quark limit will approach unity.
At large $m$ finite volume effects are small, the ratio is bigger than
1 and decreasing as $m$ increases, as expected in the heavy quark approximation.
What is remarkable in our data is the fact that in the whole
mass range we were able to explore, and in which the pseudoscalar mass changes roughly 
by a factor of 7, the vector meson never becomes more than 5\% heavier than the pseudoscalar, 
so that the ratio remains approximately constant in the chiral limit.
This is the expected behavior in an IR conformal theory, since in this
case all the hadronic masses scale with the same critical exponent.

\begin{figure}
\includegraphics*[width=\columnwidth]{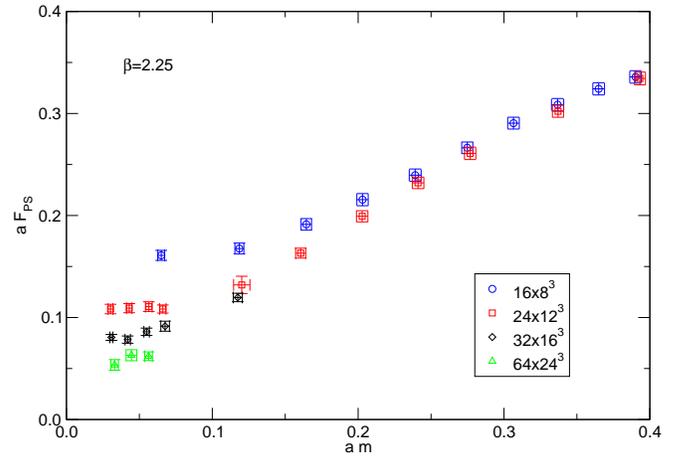}%
\caption{Pseudoscalar decay constant near the chiral limit. Very large
  finite volume effects are present also in this case which cause the
  chiral extraplation to be have large uncertainties.\label{fig:fps}}
\end{figure}
\begin{figure*}
\includegraphics*[width=1.8\columnwidth]{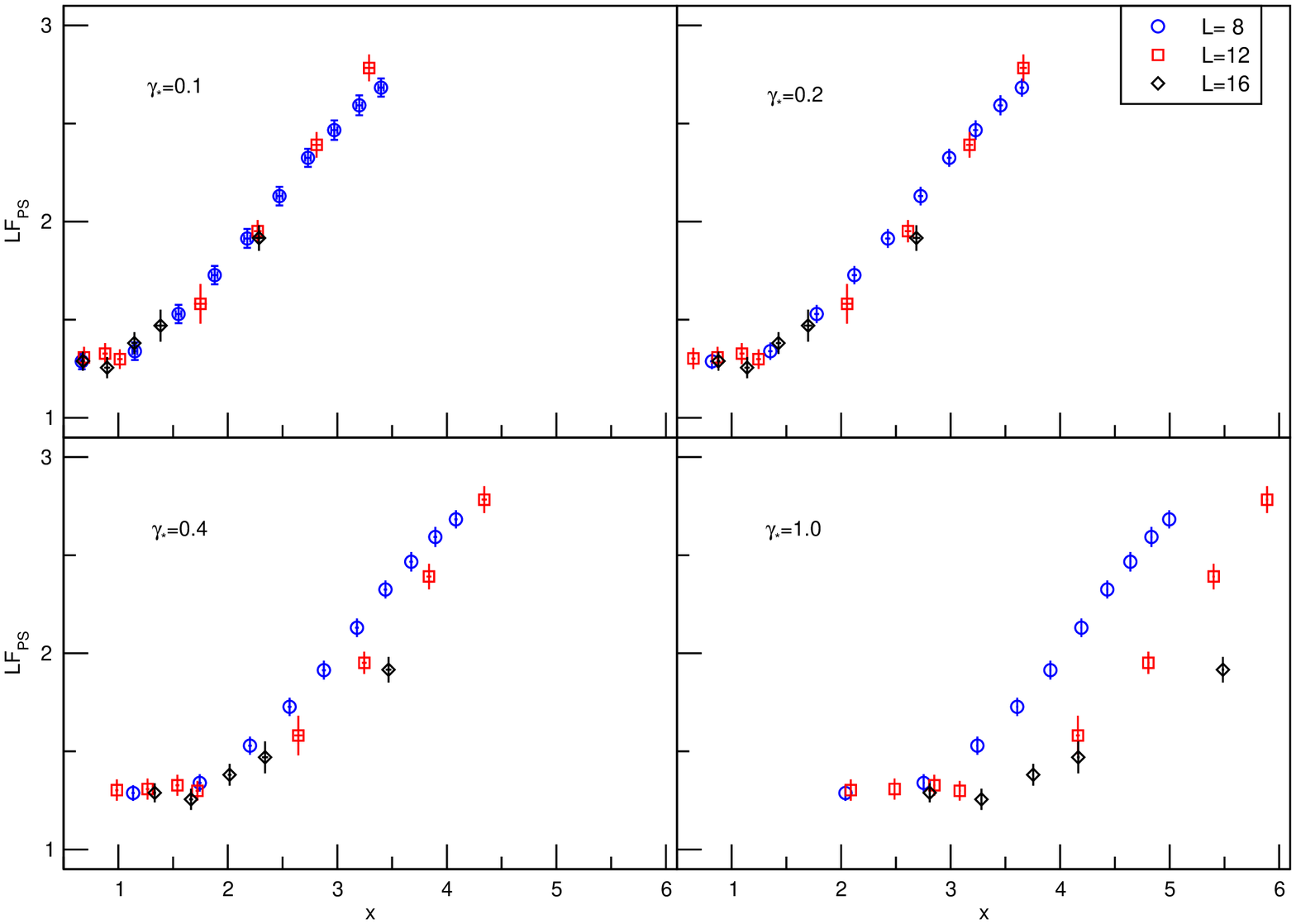}%
\caption{Quality of the scaling of $L F_{PS}$ as a function of
$x = L m^{1/(1+\gamma_*)}$ for various values of $\gamma_*$.
\label{fig:scalingfpi}}
\end{figure*}
Another physically interesting quantity to consider is the pseudoscalar decay
constant $F_\mathrm{PS}$, shown in Fig.~\ref{fig:fps}. Among the ones presented
in this paper, this is the quantity which shows the largest sensitivity to
finite-volume effects.  By looking at the behavior of $F_\mathrm{PS}$ at
different volumes, an envelope of the curves as a function of the PCAC mass is
clearly visible, which should be used for the chiral extrapolation. For a
QCD-like theory the result in the chiral limit is a non-zero value. The direct
extrapolation however is difficult to carry out with reasonable accuracy; for
this reason, we prefer to exploit the finite-size effects themselves to obtain a
more insightful statement.  As discussed in Sect.~\ref{sec:RGflow} near an IR
fixed point one can consider the finite size $L$ of the system as a relevant
parameter in the RG flux and thus obtain universal scaling laws for physical
observables, Eq.~(\ref{eq:RGfs4}). 
This finite size-scaling law can be conveniently rewritten, for example for the pseudoscalar
decay constant, as: 
\begin{equation}
L F_\mathrm{PS} = \Upsilon( L m^{1/(1 + \gamma_*)} )\, .
\end{equation}
Scaling is observed if the different curves corresponding to keeping the volumes fixed and varying the quark mass, collapse on top of each other. As a byproduct of the
procedure an estimate of the critical exponent is also obtained. To illustrate
the procedure, we plot $L F_\mathrm{PS}$ as a function of $x = L m^{1/(1 + \gamma_*)}$
in Fig.~\ref{fig:scalingfpi} for various values of $\gamma_*$. Good scaling
is observed for $0.05 \le \gamma_* \le 0.20$, while larger values of $\gamma_*$
(and in particular $\gamma_* = 1$) seem to be excluded. The observed scaling
is again in agreement with the existence of an IR fixed point with a small
$\gamma_*$ in the MWT theory. The range of values of $\gamma_*$ for which
a good quality of the scaling is obtained is compatible with independent
estimates performed with the Schr\"odinger functional~\cite{Bursa:2009we}. 

\begin{figure}
\includegraphics*[width=\columnwidth]{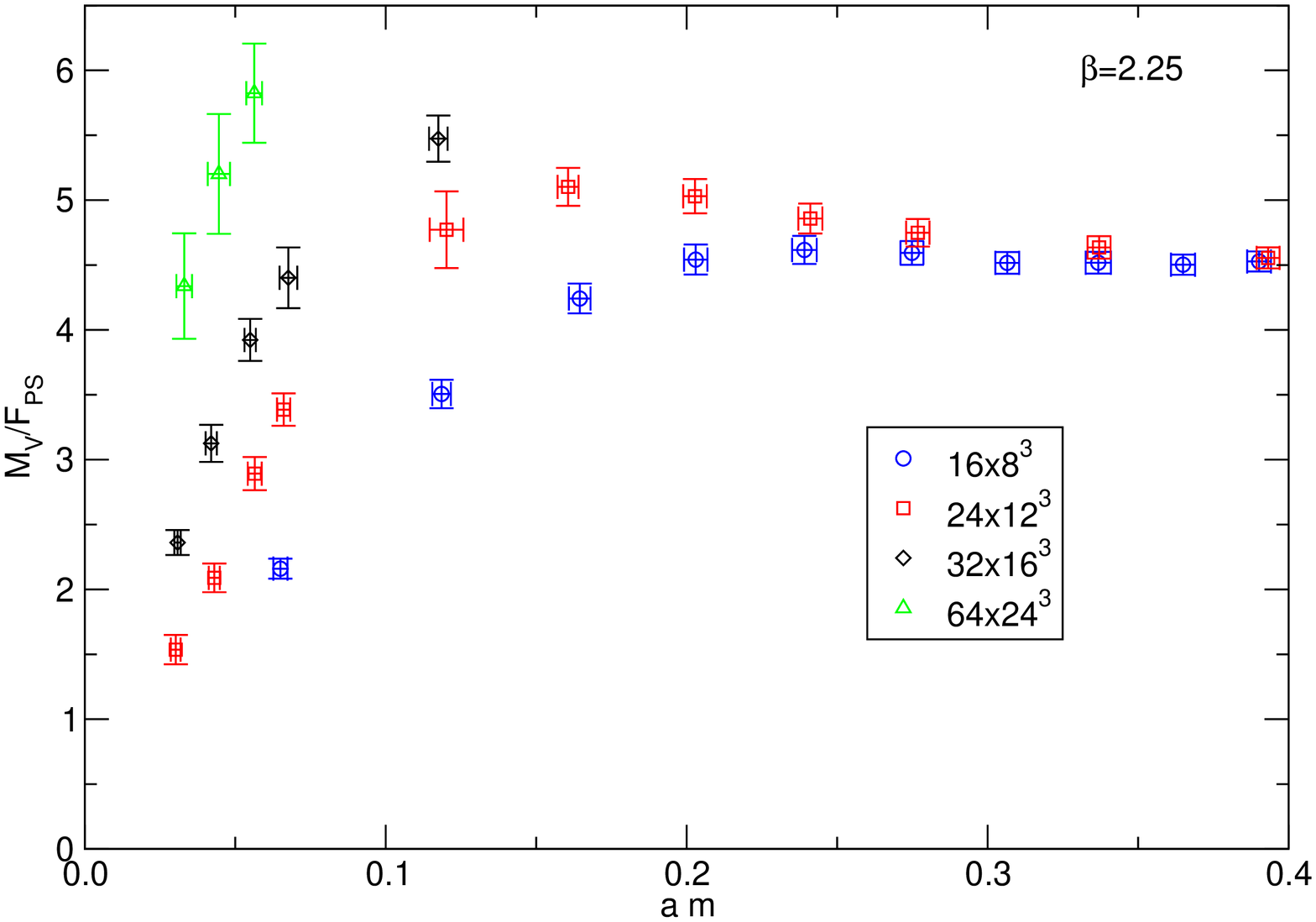}%
\caption{Vector to pseudoscalar decay constant ratio. Large finite-size
effects are present also in this case which make the extrapolation to the
chiral limit difficult. The envelop of the curves in the plot suggests a
limit value of about $5-6$.  \label{fig:mvovfpi}}
\end{figure}

A phenomenologically relevant quantity to look at is the ratio
$M_V/F_\mathrm{PS}$.  This ratio is shown in
Fig.~\ref{fig:mvovfpi}. The large dependence of $F_\mathrm{PS}$ on the
finite-size of the lattice is reflected in the large finite-size
effects for the ratio. A tentative large-volume limit curve can be
obtained by discarding the results at the lightest masses on the
smaller volumes. As the fermion mass drops below $0.1a^{-1}$, the
ratio starts decreasing unless the volume is made larger. Taking the
envelope of the curves for different values of the volume, one can
expect a value of about $5-6$ in the chiral limit.

\begin{figure}
\includegraphics*[width=\columnwidth]{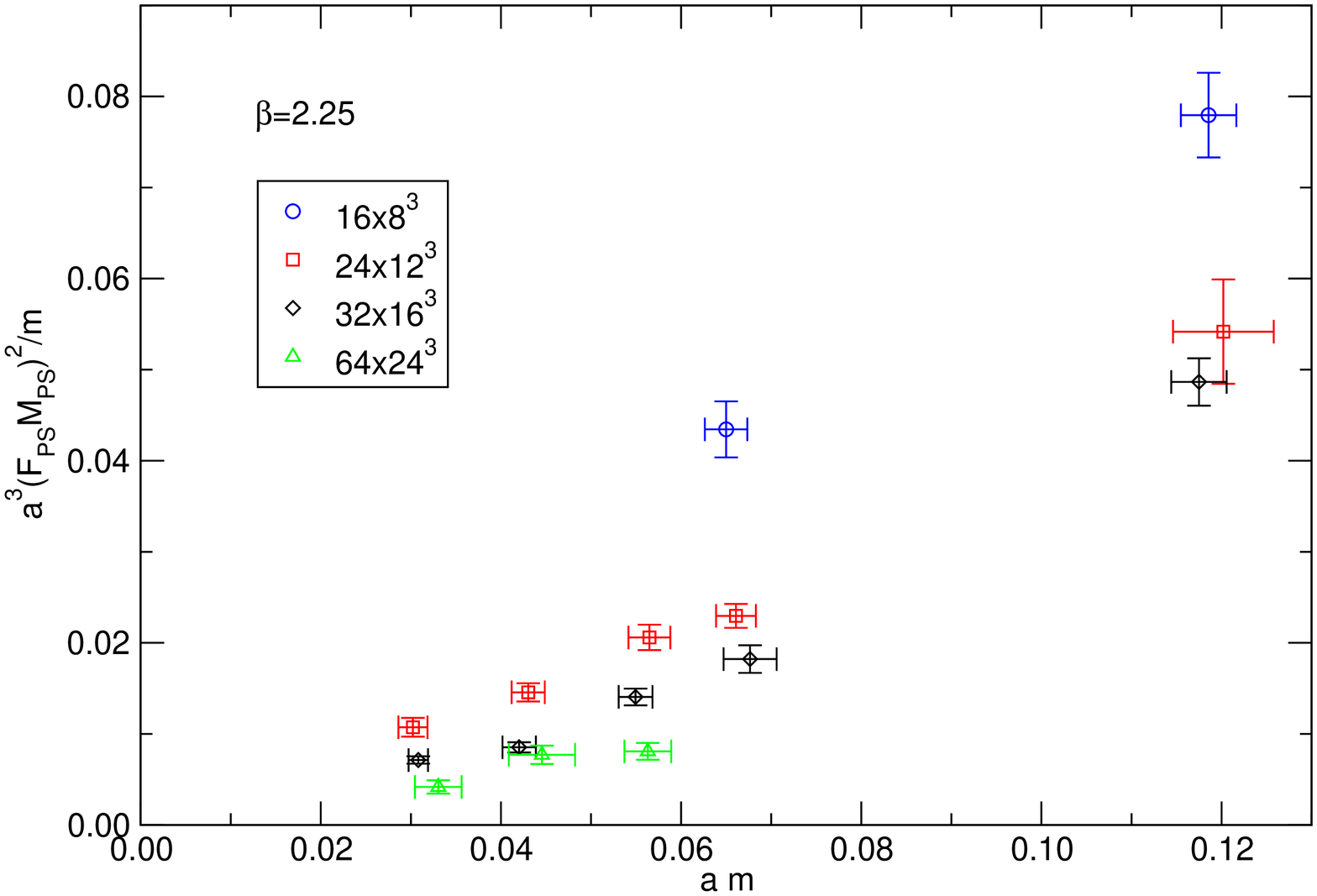}%
\caption{The GMOR relation can be used to extract information on the
  chiral condensate. The measure results however quite difficult in
  practice and we cannot distinguish any signal of spontaneous chiral symmetry
  breaking.\label{fig:gmor}}
\end{figure}
The chiral condensate would also be a prime candidate to study chiral
symmetry breaking. However due to the use of Wilson fermions, the
direct measure of $\langle\overline\psi\psi\rangle$ is plagued with UV
divergences which are notoriously difficult to tame.  Using the GMOR
relation an estimate for the chiral condensate can be
obtained\footnote{We do not attempt here to compute the necessary
  multiplicative renormalization constant, since we are not interested
  to the actual physical value. Perturbative results for the
  renormalization of fermions bilinears can be found in
  Ref.~\cite{DelDebbio:2008wb}}. The method has been applied with
success in the case of QCD, see e.g. Ref.~\cite{Giusti:1998wy}.  We
present our results for this quantity in Fig.~\ref{fig:gmor}.  Although
there is a partial cancellation of the finite size effects coming from
the pseudoscalar mass and the decay constant, the larger volume
dependence of the latter dominates, yielding large systematic errors.
As a consequence an extrapolation is unfortunately not possible from
our current set of data. We observe that finite volume effects
tend to make the condensate smaller, however the small numerical value
of the bare condesate by itself is not meaningful: for example in a typical QCD simulation
the value for this quantity is an order of magnitude smaller than the one
presented here.

\section{Conclusions}
\label{sec:concl}
In this work we have presented a careful investigation of the mesonic
spectrum of one of the candidate theories for a realistic technicolor
model, the so-called Minimal Walking Technicolor, based on gauge group
SU(2) with two Dirac adjoint fermions.  Theoretical speculations about
this theory indicate that it is very near to the lower boundary of the
conformal window.  In this work we used numerical lattice simulations
to look at mesonic spectrum and we found some evidence that the theory
lies in fact inside the conformal window and possesses an IR conformal
fixed point.

Such numerical simulations are an extremely powerful tool to explore the
non-perturbative dynamics of gauge theories which is otherwise inaccessible to
theoretical speculations, but great care must be taken to control systematic
errors.  In order  to tame finite size corrections, which make the
extrapolation to the chiral limit difficult, in this work we aimed for the
first time at reaching the chiral limit in a controlled way: we used a series
of four different lattice sizes up to a large $64\times 24^3$.

Evidence for the existence of an IR fixed point was found 
in the behavior of the different mesonic observables analyzed, namely the pseudoscalar and
vector meson mass and the pseudoscalar decay constant, which show
significant deviations from the expectations of a more familiar QCD-like
scenario, where spontaneous chiral symmetry breaking occurs.  
We showed that our data are compatible with the existence of an IR fixed point
by using the predicted scaling laws that need to hold in this case.

Although the present data show clear signs of conformality in the infrared, our
study
still has several limitations which should be addressed in the future to put
our results on a more solid ground. Smaller quark masses and consequently larger
lattice volumes would increase the reliability of the scaling analysis we
performed in this paper. However the major source of uncertainty is the fact
that all numerical simulations used in this work were performed at a single
value of the lattice spacing, and no test to assure the validity of our
findings in the continuum limit has been done so far. 

Finally in this paper we focused our attention only on the mesonic spectrum, 
while substantially more information can be gained by combining it with
observables from other sectors of the theory, as we proposed in
Ref.~\cite{DelDebbio:2009fd}. The detailed study of gluonic observables, and
their comparison to the mesonic ones 
is the subject of a companion paper~\cite{DelDebbio:2010xxx}, which provides further
evidence for the existence of an IR fixed point. 

\begin{acknowledgments}
The numerical calculations presented in this work have been performed
on the BlueC supercomputer at Swansea university, on a Beowulf cluster
partly funded by the Royal Society and on the Horseshoe5 cluster
at the supercomputing facility at the University of Southern Denmark (SDU) 
funded by a grant of the Danish Centre for Scientific Computing for the project 
``Origin of Mass" 2008/2009.
We thank C. Allton, J. Cardy, F. Knechtli, C. McNeile, M. Piai and F. Sannino for 
useful and fruitful discussions about various aspects related to this paper. 
We thank the organizers and participants of the workshop ``Universe in a box", Lorentz Center, Leiden, NL, August 2009, where some results contained in this paper were firstly presented and discussed. A.P. thanks the groups at CERN, Columbia U., Maryland U., Colorado U., Washington U., LLNL, SLAC, Syracuse U. for warmily hosting him and for useful and stimulating discussions about several aspects of this work.
Our work has been partially supported by STFC under contracts PP/E007228/1,
ST/G000506/1. B.L. is supported by the Royal Society, A.P. is supported by STFC.
A.R. thanks the Deutsche Forschungsgemeinschaft for financial support.
\end{acknowledgments}

\appendix

\section{Effective mass definition}
\label{sec:effmass}
For the definition of the effective mass used in this work we follow Ref.~\cite{Fleming:2009wb}. A mesonic correlator on the lattice has the form:
\begin{equation}
C(\tau)=\sum_{m=1}^M a_m \cosh\left[ E_m \tau \right]\,\,\, ,
\end{equation}
with $\tau=t-T/2=0,1,2,\ldots,T/2$, where we consider only $M$ excited states. Now since:
\begin{equation}
(\cosh\left[ E_m \right])^n = \frac{1}{2^n} \sum_{k=0}^n \binom{n}{k} \cosh\left[ E_m (2k-n) \right]\,\, ,
\end{equation}
taking similar linear combinations of the $C(\tau)$ we have:
\begin{equation}
\frac{1}{2^n} \sum_{k=0}^n \binom{n}{k} C(2k-n) = \sum_{m=1}^M a_m (\cosh\left[ E_m \right])^n \,\,\, .
\label{VMsystem1}
\end{equation}
Introducing the variables:
\begin{eqnarray}
x_m&\equiv&\cosh\left[ E_m \right]\,\, ,\\
y_n&\equiv& \frac{1}{2^n} \sum_{k=0}^n \binom{n}{k} C(2k-n)\,\, ,
\end{eqnarray}
we can rewrite Eq.~\ref{VMsystem1} in matrix form as:
\begin{equation}
\begin{pmatrix}
y_0\\
y_1\\
\vdots
\end{pmatrix}=
\begin{pmatrix}
1&1&1&1&\cdots\\
x_1&x_2&x_3&x_4&\cdots\\
x_1^2&x_2^2&x_3^2&x_4^2&\cdots\\
x_1^3&x_2^3&x_3^3&x_4^3&\cdots\\
\vdots
\end{pmatrix}\cdot
\begin{pmatrix}
a_1\\
a_2\\
\vdots
\end{pmatrix}\,\,\, .
\label{VMsystem2}
\end{equation}
We need to solve Eq.~\ref{VMsystem2} where $y_n$ is known and both $a_m$ and $x_m$ are unknown. It is always possible to find the unique solution to Eq.~\ref{VMsystem2} considering $2M$ consecutive points in the transformed correlator $y_n$. In general the $x_n$ are given by the roots of the $M$-degree polynomial:
\begin{equation}
\det \begin{pmatrix}
y_0&y_1&\cdots&y_{M-1}&1\\
y_1&y_2&\cdots&y_{M}&x\\
y_2&y_3&\cdots&y_{M+1}&x^2\\
\vdots&\vdots&&\vdots&\vdots\\
y_M&y_{M+1}&\cdots&y_{2M-1}&x^M\\
\end{pmatrix}=0\,\,\,\, ,
\label{VMsol}
\end{equation}
and the $a_m$ are then given by the solution of the linear system Eq.(4) obtained with the known $x_n$.

In this work we do not consider excited states and we only need the solution of Eq.~\ref{VMsol} for $M=1$ which is given by $x = y_1/y_0$.


\section{Tables}
\label{sec:tables}

\
\begin{table*}[h]
\begin{ruledtabular}
\include{TABLES/tbl_16x8x8x8b2.25_stat}
\caption{Bare parameters and average plaquette for the $16 \times 8^3$ lattice.\label{TBL1_A}}
\end{ruledtabular}
\end{table*}

\
\begin{table*}[h]
\begin{ruledtabular}
\include{TABLES/tbl_16x8x8x8b2.25_res2.cmb}
\caption{PCAC and meson masses from the $16 \times 8^3$ lattice.\label{TBL2_A}}
\end{ruledtabular}
\end{table*}

\
\begin{table*}[h]
\begin{ruledtabular}
\include{TABLES/tbl_16x8x8x8b2.25_ratio_res2.cmb}
\caption{Mass ratios from the $16 \times 8^3$ lattice.\label{TBL3_A}}
\end{ruledtabular}
\end{table*}

\
\begin{table*}[h]
\begin{ruledtabular}
\include{TABLES/tbl_24x12x12x12b2.25_stat}
\caption{Bare parameters and average plaquette for the $24 \times 12^3$ lattice.\label{TBL1_B}}
\end{ruledtabular}
\end{table*}

\
\begin{table*}[h]
\begin{ruledtabular}
\include{TABLES/tbl_24x12x12x12b2.25_res2.cmb}
\caption{PCAC and meson masses from the $24 \times 12^3$ lattice.\label{TBL2_B}}
\end{ruledtabular}
\end{table*}

\
\begin{table*}[h]
\begin{ruledtabular}
\include{TABLES/tbl_24x12x12x12b2.25_ratio_res2.cmb}
\caption{Mass ratios from the $24 \times 12^3$ lattice.\label{TBL3_B}}
\end{ruledtabular}
\end{table*}

\
\begin{table*}[h]
\begin{ruledtabular}
\include{TABLES/tbl_32x16x16x16b2.25_stat}
\caption{Bare parameters and average plaquette for the $32 \times 16^3$ lattice.\label{TBL1_C}}
\end{ruledtabular}
\end{table*}

\
\begin{table*}[h]
\begin{ruledtabular}
\include{TABLES/tbl_32x16x16x16b2.25_res2.cmb}
\caption{PCAC and meson masses from the $32 \times 16^3$ lattice.\label{TBL2_C}}
\end{ruledtabular}
\end{table*}

\
\begin{table*}[h]
\begin{ruledtabular}
\include{TABLES/tbl_32x16x16x16b2.25_ratio_res2.cmb}
\caption{Mass ratios from the $32 \times 16^3$ lattice.\label{TBL3_C}}
\end{ruledtabular}
\end{table*}

\
\begin{table*}[h]
\begin{ruledtabular}
\include{TABLES/tbl_64x24x24x24b2.25_stat}
\caption{Bare parameters and average plaquette for the $64 \times 24^3$ lattice.\label{TBL1_D}}
\end{ruledtabular}
\end{table*}

\
\begin{table*}[h]
\begin{ruledtabular}
\include{TABLES/tbl_64x24x24x24b2.25_res2.cmb}
\caption{PCAC and meson masses from the $64 \times 24^3$ lattice.\label{TBL2_D}}
\end{ruledtabular}
\end{table*}

\
\begin{table*}[h]
\begin{ruledtabular}
\include{TABLES/tbl_64x24x24x24b2.25_ratio_res2.cmb}
\caption{Mass ratios from the $64 \times 24^3$ lattice.\label{TBL3_D}}
\end{ruledtabular}
\end{table*}

\bibliography{mesons}

\end{document}

%% file: TABLES/tbl_16x8x8x8b2.25_stat.tex
\begin{tabular}{cccccccc}
lattice & V & $-am_0$ & $N_{traj}$ & $\langle P\rangle$ & $\tau$ & $\lambda$ & $\tau_\lambda$ \\
\hline
A0 & $16\times 8^3$ & 0.95 & 7601 & 0.63577(16) & 5.45(72) & 3.582(13) & 8.6(1.4)  \\
A1 & $16\times 8^3$ & 0.975 & 7701 & 0.63843(15) & 5.43(71) & 2.982(12) & 6.65(96)  \\
A2 & $16\times 8^3$ & 1 & 7801 & 0.64136(15) & 5.10(64) & 2.427(11) & 6.28(88)  \\
A3 & $16\times 8^3$ & 1.025 & 7801 & 0.64463(15) & 4.29(50) & 1.894(10) & 6.07(84)  \\
A4 & $16\times 8^3$ & 1.05 & 7801 & 0.64793(15) & 3.48(36) & 1.4596(79) & 4.39(52)  \\
A5 & $16\times 8^3$ & 1.075 & 6400 & 0.65179(16) & 2.99(32) & 1.0692(74) & 4.27(55)  \\
A6 & $16\times 8^3$ & 1.1 & 6400 & 0.65566(16) & 3.28(37) & 0.7564(60) & 3.77(45)  \\
A7 & $16\times 8^3$ & 1.125 & 7073 & 0.66037(15) & 2.99(30) & 0.4854(43) & 3.03(31)  \\
A8 & $16\times 8^3$ & 1.15 & 6400 & 0.66550(16) & 3.31(37) & 0.2779(31) & 2.80(29)  \\
A9 & $16\times 8^3$ & 1.175 & 6400 & 0.67177(17) & 3.24(36) & 0.1351(18) & 2.80(29)  \\
\end{tabular}

%% file: TABLES/tbl_16x8x8x8b2.25_res2.cmb.tex
\begin{tabular}{ccccccc}
lattice & $-am_0$ & $am$ & $aM_\mathrm{PS}$ & $aM_\mathrm{V}$ & $aF_{PS}$ & $a^2 G_{PS}$\\
\hline
A0 & 0.95 & 0.3899(40) & 1.4717(40) & 1.5203(52) & 0.3354(60) & 0.933(13)  \\
A1 & 0.975 & 0.3649(41) & 1.4093(43) & 1.4586(55) & 0.3240(61) & 0.883(13)  \\
A2 & 1 & 0.3365(42) & 1.3436(43) & 1.3936(58) & 0.3083(62) & 0.829(13)  \\
A3 & 1.025 & 0.3066(38) & 1.2630(46) & 1.3115(59) & 0.2908(58) & 0.759(12)  \\
A4 & 1.05 & 0.2749(39) & 1.1756(48) & 1.2233(64) & 0.2666(58) & 0.673(11)  \\
A5 & 1.075 & 0.2389(39) & 1.0623(56) & 1.1048(72) & 0.2392(59) & 0.567(11)  \\
A6 & 1.1 & 0.2031(39) & 0.9398(66) & 0.9784(86) & 0.2157(59) & 0.472(11)  \\
A7 & 1.125 & 0.1643(36) & 0.7817(76) & 0.811(10) & 0.1910(57) & 0.357(10)  \\
A8 & 1.15 & 0.1185(32) & 0.5740(89) & 0.587(11) & 0.1675(56) & 0.2347(82)  \\
A9 & 1.175 & 0.0650(24) & 0.330(11) & 0.3476(91) & 0.1611(50) & 0.1347(76)  \\
\end{tabular}

%% file: TABLES/tbl_16x8x8x8b2.25_ratio_res2.cmb.tex
\begin{tabular}{ccccccc}
lattice & $-am_0$ & $am$ & $aM_\mathrm{PS}^2/m$ & $M_\mathrm{V}/F_{PS}$ & $M_\mathrm{V}/M_\mathrm{PS}$ & $a^3 (M_\mathrm{PS}F_{PS})^2/m$\\
\hline
A0 & 0.95 & 0.3899(40) & 5.554(62) & 4.533(77) & 1.0330(12) & 0.625(19)  \\
A1 & 0.975 & 0.3649(41) & 5.443(67) & 4.502(79) & 1.0349(13) & 0.571(18)  \\
A2 & 1 & 0.3365(42) & 5.365(71) & 4.521(86) & 1.0372(15) & 0.510(17)  \\
A3 & 1.025 & 0.3066(38) & 5.203(70) & 4.511(84) & 1.0384(17) & 0.440(15)  \\
A4 & 1.05 & 0.2749(39) & 5.027(77) & 4.589(94) & 1.0405(22) & 0.357(13)  \\
A5 & 1.075 & 0.2389(39) & 4.724(83) & 4.62(10) & 1.0399(28) & 0.270(11)  \\
A6 & 1.1 & 0.2031(39) & 4.349(92) & 4.53(11) & 1.0410(37) & 0.2025(96)  \\
A7 & 1.125 & 0.1643(36) & 3.721(92) & 4.24(11) & 1.0375(53) & 0.1358(74)  \\
A8 & 1.15 & 0.1185(32) & 2.782(90) & 3.51(11) & 1.0242(88) & 0.0781(48)  \\
A9 & 1.175 & 0.0650(24) & 1.67(10) & 2.159(80) & 1.054(31) & 0.0434(30)  \\
\end{tabular}

%% file: TABLES/tbl_24x12x12x12b2.25_stat.tex
\begin{tabular}{cccccccc}
lattice & V & $-am_0$ & $N_{traj}$ & $\langle P\rangle$ & $\tau$ & $\lambda$ & $\tau_\lambda$ \\
\hline
B0 & $24\times 12^3$ & 0.95 & 10201 & 0.635310(59) & 6.16(74) & 3.5058(50) & 3.08(26)  \\
B1 & $24\times 12^3$ & 1 & 8652 & 0.640998(64) & 4.92(58) & 2.4218(44) & 3.10(29)  \\
B2 & $24\times 12^3$ & 1.05 & 7819 & 0.647633(70) & 6.79(99) & 1.4936(51) & 5.80(78)  \\
B3 & $24\times 12^3$ & 1.075 & 7186 & 0.651630(68) & 4.61(58) & 1.0553(40) & 4.95(64)  \\
B4 & $24\times 12^3$ & 1.1 & 6393 & 0.655827(76) & 4.09(51) & 0.7202(30) & 7.8(1.3)  \\
B5 & $24\times 12^3$ & 1.125 & 6200 & 0.660588(75) & 3.97(50) & 0.4419(22) & 5.98(91)  \\
B6 & $24\times 12^3$ & 1.15 & 1599 & 0.66588(15) & 3.71(90) & 0.2271(31) & 6.6(2.1)  \\
B7 & $24\times 12^3$ & 1.175 & 5582 & 0.672074(79) & 4.22(58) & 0.08641(90) & 3.78(49)  \\
B8 & $24\times 12^3$ & 1.18 & 4081 & 0.673474(92) & 4.01(63) & 0.06561(92) & 10(2.5)  \\
B9 & $24\times 12^3$ & 1.185 & 4201 & 0.675094(93) & 3.42(49) & 0.05196(71) & 3.53(51)  \\
B10 & $24\times 12^3$ & 1.19 & 3501 & 0.67663(10) & 4.15(70) & 0.03985(61) & 5.2(1.0)  \\
\end{tabular}

%% file: TABLES/tbl_24x12x12x12b2.25_res2.cmb.tex
\begin{tabular}{ccccccc}
lattice & $-am_0$ & $am$ & $aM_\mathrm{PS}$ & $aM_\mathrm{V}$ & $aF_{PS}$ & $a^2 G_{PS}$\\
\hline
B0 & 0.95 & 0.3931(38) & 1.4746(23) & 1.5224(32) & 0.3343(62) & 0.925(12)  \\
B1 & 1 & 0.3368(40) & 1.3495(26) & 1.4003(36) & 0.3020(63) & 0.819(12)  \\
B2 & 1.05 & 0.2765(40) & 1.1874(29) & 1.2383(40) & 0.2607(63) & 0.667(11)  \\
B3 & 1.075 & 0.2410(38) & 1.0809(30) & 1.1265(41) & 0.2320(58) & 0.5635(96)  \\
B4 & 1.1 & 0.2025(40) & 0.9614(35) & 1.0016(46) & 0.1991(53) & 0.4558(85)  \\
B5 & 1.125 & 0.1604(34) & 0.8020(41) & 0.8312(56) & 0.1628(49) & 0.3277(71)  \\
B6 & 1.15 & 0.1198(52) & 0.6111(91) & 0.627(11) & 0.1313(78) & 0.2066(97)  \\
B7 & 1.175 & 0.0660(22) & 0.3593(52) & 0.3659(66) & 0.1083(40) & 0.1055(33)  \\
B8 & 1.18 & 0.0565(23) & 0.3085(60) & 0.3199(76) & 0.1108(47) & 0.0927(35)  \\
B9 & 1.185 & 0.0430(18) & 0.2292(69) & 0.2277(85) & 0.1090(44) & 0.0664(32)  \\
B10 & 1.19 & 0.0302(16) & 0.1664(81) & 0.165(10) & 0.1083(45) & 0.0506(34)  \\
\end{tabular}

%% file: TABLES/tbl_24x12x12x12b2.25_ratio_res2.cmb.tex
\begin{tabular}{ccccccc}
lattice & $-am_0$ & $am$ & $aM_\mathrm{PS}^2/m$ & $M_\mathrm{V}/F_{PS}$ & $M_\mathrm{V}/M_\mathrm{PS}$ & $a^3 (M_\mathrm{PS}F_{PS})^2/m$\\
\hline
B0 & 0.95 & 0.3931(38) & 5.531(55) & 4.555(81) & 1.03241(85) & 0.618(19)  \\
B1 & 1 & 0.3368(40) & 5.406(67) & 4.637(93) & 1.0376(11) & 0.493(17)  \\
B2 & 1.05 & 0.2765(40) & 5.098(77) & 4.75(11) & 1.0428(14) & 0.346(13)  \\
B3 & 1.075 & 0.2410(38) & 4.849(78) & 4.85(11) & 1.0421(16) & 0.261(10)  \\
B4 & 1.1 & 0.2025(40) & 4.564(93) & 5.03(12) & 1.0418(20) & 0.1810(75)  \\
B5 & 1.125 & 0.1604(34) & 4.010(90) & 5.10(14) & 1.0364(30) & 0.1063(51)  \\
B6 & 1.15 & 0.1198(52) & 3.12(15) & 4.79(27) & 1.0272(80) & 0.0539(53)  \\
B7 & 1.175 & 0.0660(22) & 1.958(74) & 3.38(12) & 1.0183(86) & 0.0229(13)  \\
B8 & 1.18 & 0.0565(23) & 1.687(81) & 2.89(13) & 1.036(13) & 0.0207(13)  \\
B9 & 1.185 & 0.0430(18) & 1.223(73) & 2.09(11) & 0.993(18) & 0.0145(10)  \\
B10 & 1.19 & 0.0302(16) & 0.918(85) & 1.53(11) & 0.996(39) & 0.01076(98)  \\
\end{tabular}

%% file: TABLES/tbl_32x16x16x16b2.25_stat.tex
\begin{tabular}{cccccccc}
lattice & V & $-am_0$ & $N_{traj}$ & $\langle P\rangle$ & $\tau$ & $\lambda$ & $\tau_\lambda$ \\
\hline
C0 & $32\times 16^3$ & 1.15 & 5446 & 0.665894(44) & 3.32(40) & 0.2227(10) & 3.05(36)  \\
C1 & $32\times 16^3$ & 1.175 & 2192 & 0.672235(73) & 2.80(50) & 0.07036(90) & 5.9(1.5)  \\
C2 & $32\times 16^3$ & 1.18 & 4606 & 0.673657(49) & 3.46(47) & 0.05167(50) & 6.1(1.1)  \\
C3 & $32\times 16^3$ & 1.185 & 4313 & 0.675170(50) & 2.99(39) & 0.03751(38) & 4.66(75)  \\
C4 & $32\times 16^3$ & 1.19 & 5404 & 0.676637(44) & 3.29(40) & 0.02474(28) & 7.9(1.5)  \\
\end{tabular}

%% file: TABLES/tbl_32x16x16x16b2.25_res2.cmb.tex
\begin{tabular}{ccccccc}
lattice & $-am_0$ & $am$ & $aM_\mathrm{PS}$ & $aM_\mathrm{V}$ & $aF_{PS}$ & $a^2 G_{PS}$\\
\hline
C0 & 1.15 & 0.1175(30) & 0.6319(31) & 0.6541(43) & 0.1196(41) & 0.2037(49)  \\
C1 & 1.175 & 0.0678(30) & 0.3834(49) & 0.4015(61) & 0.0919(51) & 0.1018(37)  \\
C2 & 1.18 & 0.0549(18) & 0.3226(37) & 0.3364(46) & 0.0860(35) & 0.0817(24)  \\
C3 & 1.185 & 0.0420(16) & 0.2416(39) & 0.2443(50) & 0.0784(32) & 0.0542(18)  \\
C4 & 1.19 & 0.0308(10) & 0.1842(36) & 0.1900(43) & 0.0806(29) & 0.0443(15)  \\
\end{tabular}

%% file: TABLES/tbl_32x16x16x16b2.25_ratio_res2.cmb.tex
\begin{tabular}{ccccccc}
lattice & $-am_0$ & $am$ & $aM_\mathrm{PS}^2/m$ & $M_\mathrm{V}/F_{PS}$ & $M_\mathrm{V}/M_\mathrm{PS}$ & $a^3 (M_\mathrm{PS}F_{PS})^2/m$\\
\hline
C0 & 1.15 & 0.1175(30) & 3.398(90) & 5.47(18) & 1.0351(35) & 0.0486(26)  \\
C1 & 1.175 & 0.0678(30) & 2.17(10) & 4.38(23) & 1.0473(93) & 0.0183(15)  \\
C2 & 1.18 & 0.0549(18) & 1.895(70) & 3.91(16) & 1.0426(86) & 0.01404(89)  \\
C3 & 1.185 & 0.0420(16) & 1.390(61) & 3.12(13) & 1.011(12) & 0.00854(52)  \\
C4 & 1.19 & 0.0308(10) & 1.102(53) & 2.36(10) & 1.031(18) & 0.00715(38)  \\
\end{tabular}

%% file: TABLES/tbl_64x24x24x24b2.25_stat.tex
\begin{tabular}{cccccccc}
lattice & V & $-am_0$ & $N_{traj}$ & $\langle P\rangle$ & $\tau$ & $\lambda$ & $\tau_\lambda$ \\
\hline
D0 & $64\times 24^3$ & 1.18 & 458 & 0.673737(46) & 4.0(1.9) & 0.04436(51) & 3.5(1.5)  \\
D1 & $64\times 24^3$ & 1.185 & 291 & 0.675184(59) & 2.3(1.1) & 0.02836(59) & 4.2(2.5)  \\
D2 & $64\times 24^3$ & 1.19 & 349 & 0.676649(52) & 1.63(59) & 0.01520(39) & 5.7(3.6)  \\
\end{tabular}

%% file: TABLES/tbl_64x24x24x24b2.25_res2.cmb.tex
\begin{tabular}{ccccccc}
lattice & $-am_0$ & $am$ & $aM_\mathrm{PS}$ & $aM_\mathrm{V}$ & $aF_{PS}$ & $a^2 G_{PS}$\\
\hline
D0 & 1.18 & 0.0562(25) & 0.3433(37) & 0.3597(56) & 0.0620(42) & 0.0661(31)  \\
D1 & 1.185 & 0.0445(36) & 0.2930(73) & 0.325(11) & 0.0629(53) & 0.0580(52)  \\
D2 & 1.19 & 0.0330(25) & 0.2184(73) & 0.230(10) & 0.0534(50) & 0.0388(38)  \\
\end{tabular}

%% file: TABLES/tbl_64x24x24x24b2.25_ratio_res2.cmb.tex
\begin{tabular}{ccccccc}
lattice & $-am_0$ & $am$ & $aM_\mathrm{PS}^2/m$ & $M_\mathrm{V}/F_{PS}$ & $M_\mathrm{V}/M_\mathrm{PS}$ & $a^3 (M_\mathrm{PS}F_{PS})^2/m$\\
\hline
D0 & 1.18 & 0.0562(25) & 2.099(97) & 5.82(37) & 1.0477(98) & 0.00808(93)  \\
D1 & 1.185 & 0.0445(36) & 1.94(19) & 5.20(46) & 1.110(22) & 0.00768(99)  \\
D2 & 1.19 & 0.0330(25) & 1.45(13) & 4.34(40) & 1.054(24) & 0.00415(73)  \\
\end{tabular}